\documentclass[twocolumn,aps,prd,amssymb,eqsecnum,nofootinbib,superscriptaddress,showpacs]{revtex4}
\usepackage{ifpdf} \ifpdf \usepackage{hyperref} \else
\usepackage[dvipdfm]{hyperref} \fi

\usepackage{amsmath} \usepackage{amssymb} \usepackage{epsfig}
\usepackage{graphicx}
\usepackage{psfrag} \usepackage{rotating} \usepackage{multirow}
\usepackage[percent]{overpic}
\usepackage{overpic}
\usepackage{bm}

\newcommand{\be}{\begin{equation}}
\newcommand{\ee}{\end{equation}}
\newcommand{\ba}{\begin{eqnarray}}
\newcommand{\ea}{\end{eqnarray}}
\newcommand{\bma}{\begin{pmatrix}}
\newcommand{\ema}{\end{pmatrix}}
\newcommand{\bea}{\begin{eqnarray}}
\newcommand{\eea}{\end{eqnarray}}

\newcommand{\WVU}{\affiliation{Department of Physics, West Virginia University, PO Box 6315, Morgantown,
West Virginia 26506}}

\begin{document}
\title{Decline of the current quadrupole moment during the merger phase \\
of binary black hole coalescence}

\author{Fan Zhang} \WVU
\date{printed \today}
\begin{abstract} 
Utilizing the tools of tendex and vortex, we study the highly dynamic plunge and merger phases of several $\pi$-symmetric binary black hole coalescences. 
In particular, we observe a decline of the strength of the current quadrupole moment as compared to that of the mass quadrupole moment during the merger phase, contrary to a naive estimate according to the dependence of these moments on the separation between the black holes. We further show that this decline of the current quadrupole moment is achieved through the remnants of the two individual spins becoming nearly aligned or anti-aligned with the total angular momentum. We also speculate on the implication of our observations for achieving a consistency between the electric and magnetic parity quasinormal modes. 
\end{abstract}


\maketitle  

\section{Introduction \label{sec:Intro}}
Binary black hole (BBH) coalescences constitute one of the most promising types of gravitational wave sources for the network of detectors, such as the Advanced LIGO \cite{Harry:2010zz}, Virgo \cite{aVIRGO,aVIRGO:2012}, GEO \cite{Grote:2010zz}, and KAGRA \cite{Somiya:2012}. Beyond an initial detection, gravitational wave astronomy is also on the horizon (see e.g. \cite{Camp:2004gg}), and so it is important to know what 
dynamics of the astronomical system underlies the inspiral, merger and ringdown stages of a BBH waveform, and can therefore be studied using that waveform. 
For example, the merger phase (defined here as between the formation of the common apparent horizon, i.e. merger, and the beginning of the quasinormal mode ringdown) dynamics are interesting because they reflect strong gravity behaviors and correspond to a large gravitational wave amplitude. 

\begin{table*}[!t]
\begin{tabular}{c|c|c|c|c}
  \hline
Simulation label & SK & SK- & SK$\bot$ & AA \\
\hline\hline 
\hspace{1mm} Initial ADM angular momentum \hspace{1mm} & 
\hspace{2mm} (0,0,0.9844) \hspace{2mm} & 
\hspace{2mm} (0,0,0.9825) \hspace{2mm} &
\hspace{2mm} (0,0,0.9682) \hspace{2mm} &
\hspace{2mm} (0,0,0.7475) \hspace{2mm} 
\\ \hline
Initial ADM energy & 0.9903 & 0.9903 & 0.9899 & 0.9901
\\ \hline
Initial $\Omega$ & 0.02668 & 0.02668 & 0.02668 & 0.02668 
\\ \hline 
Initial Christodoulou mass 1 & 0.5 & 0.5 & 0.5 & 0.5
\\ \hline
Initial Christodoulou mass 2 & 0.5 & 0.5 & 0.5 & 0.5 
\\ \hline
Initial Dimensionless spin 1 & (0.5,0,0) & (-0.5,0,0) & (0,0.5,0) & (0,0,-0.5)
\\ \hline
Initial Dimensionless spin 2 & (-0.5,0,0) & (0.5,0,0) & (0,-0.5,0) & (0,0,-0.5)
\\ \hline \hline
Final Christodoulou mass & 0.95 & 0.95 & 0.95 & 0.96
\\ \hline
Final Dimensionless spin & 0.68 & 0.68 & 0.68 & 0.53
\\ \hline 
\end{tabular}
\caption{Initial parameters for the BBH simulations. In all the simulations, the black holes are initially on the $x$ axis, and the orbital plane is spanned by the $x$ and $y$ axes, while the total angular momentum is in the $\bm{\hat{z}}$ direction (the hat signifies unit magnitude). We also include the mass and spin of the final remnant black holes in the bottom two rows. }
\label{tb:superkickID}
\end{table*}

The particular aspect of the merger phase dynamics we examine is the decline (not necessarily disappearance) of the current quadrupole moment relative to the mass quadrupole moment in the near zone. For our study, we will rely on $\pi$-symmetric simulations such as the superkick (equal-mass BBH with anti-aligned spins in the orbital plane) BBH coalescence previously examined in Ref.~\cite{OwenEtAl:2011} for demonstration, which we will refer to as the SK simulation. By $\pi$-symmetry, we mean that the system is invariant under a $\pi$-rotation around an axis orthogonal to the initial orbital plane \cite{Bruegmann-Gonzalez-Hannam-etal:2007}. This symmetry brings about significant simplifications that are useful for us.
We will also utilize other $\pi$-symmetric simulations (not necessarily superkick configurations) whose initial parameters are similar to those of SK, aside from the initial spin orientations.
The details of the initial parameters for these simulations are given in Table \ref{tb:superkickID}.

In this paper, weak field and/or perturbative expressions are utilized to help build intuition and aid in the formulation of qualitative arguments. However, we will use the tools of tendex and vortex, which are non-perturbative and valid in strong fields, to examine the numerical simulations. We begin by examining the analytical predictions for the tendex and vortex fields generated by the mass and current quadrupoles in Sec.~\ref{sec:Analytic}. We then use the knowledge gained to study these quantities in the SK simulation, and show that the current quadrupole declines in relative importance against the mass quadrupole during the merger phase. In Sec.~\ref{sec:ExitStrategy}, we propose the mechanism through which the current quadrupole makes its exit, namely that the remnants of the individual spins become (nearly) aligned or anti-aligned with the total angular momentum. In Sec.~\ref{sec:ObRemSpin}, we directly visualize the movements of these remnant spins using the horizon vorticity, which appear to be in agreement with our proposal. Finally in Sec.~\ref{sec:Discussion}, we speculate on the implication of our observations in terms of helping the electric and magnetic parity quasinormal modes (QNM) achieve equality in their frequencies.  

Note that the spin-total angular momentum alignment or anti-alignment considered here is not the same as the spin-flip examined in, for example, Refs.~\cite{Merritt2002,Campanelli2007b}, which considers the difference between the spin of the remnant black hole and the pre-merger individual spins (i.e. a comparison between different entities), and is simply a result of the former acquiring much of the pre-merger orbital angular momentum \cite{Campanelli2007b}. Our discussion is, however, a comparison between the individual (remnant) spins with their earlier selves. 

In the formulas below, the spacetime indices are written in the front part of the Latin alphabet, while the spatial indices use the middle part of that alphabet. We will use bold-face font for vectors and tensors, and adopt geometrized units with $c=1=G$. All the simulations and visualizations are performed with the Spectral Einstein Code (\verb!SpEC!)
\cite{SXSWebsite} infrastructure.
  
\section{Vorticity from the mass and current quadrupoles \label{sec:Analytic}}
Given a timelike vector field $\bm{u}$ normal to a foliation of the spacetime by spatial hypersurfaces, the tendex $\bm{\mathcal{E}}$ and vortex $\bm{\mathcal{B}}$ fields are spatial tensors defined by 
\bea \label{eq:DefEB}
\mathcal{E}_{ij} + i \mathcal{B}_{ij} = h_i{}^e h_j{}^f \left(C_{ecfd} - i{}^*C_{ecfd} \right) u^c u^d, 
\eea
where $C_{abcd}$ is the Weyl curvature tensor, $h_{ab}=g_{ab}+ u_a u_b$ is the projection operator into the spatial hypersurfaces with $g_{ab}$ being the spacetime metric, and the Hodge dual operates on the first two indices. 
Because the Weyl curvature tensor can be decomposed into and be reconstructed from the tendex and vortex fields, we can see these fields as representations of the spacetime geometry.   
The eigenvalues of $\bm{\mathcal{E}}$ and $\bm{\mathcal{B}}$ are called tendicities and vorticities. Because the tendex and vortex tensors are $3\times 3$ matrices at each field location, there are three branches of tendicities and vorticities. From the discussions in Sec.~VI of Ref.~\cite{Nichols:2011pu}, we know that in the wave zone, one of the branches is associated with the Coulomb background piece of the Weyl curvature tensor, in the sense that the tendicity and vorticity are the real and imaginary parts of the Newman-Penrose (NP) scalar $\Psi_2$ (see Ref.~\cite{Szekeres1965} for more details on interpreting $\Psi_2$ as the Coulomb background, and $\Psi_4$ as the outgoing transverse radiation). We will refer to this branch as the Coulomb branch, even in the near zone. The other two branches weave into the gravitational wavefront in the sense of Figs.~7 and 8 of Ref.~\cite{Nichols:2011pu}, as well as Ref.~\cite{Zimmerman2011}. Another definition we will need is the horizon vorticity. Given the spatial normal $\bm{N}$ to an apparent or event horizon, the horizon vorticity $\mathcal{B}_{NN}$ is defined by $\mathcal{B}_{NN} \equiv \mathcal{B}_{ij}N^iN^j$.  

For the rest of the section, we will mostly specialize to the Coulomb branch vorticity field generated by the mass and current quadrupoles, although as the discussions are centered on symmetries, they would work with the other two branches as well.   
The mass quadrupole contains only orbital motion contribution; and is given by 
\bea \label{eq:MassQuadrupole}
\mathcal{I}_{jk} = \left(\sum_{A} m_A x_{Aj}x_{Ak} \right)^{\text{STF}},
\eea
where $A \in \{1,2\}$ labels the black holes and $\text{STF}$ stands for taking the symmetric, trace-free part. The current quadrupole moment is given by 
\bea \label{eq:CurrQuad}
\mathcal{S}_{jk} = \left( \sum_{A} x_{Aj}J^{\rm tot}_{Ak} \right)^{\text{STF}},
\eea
with $\bm{J}^{\rm tot}$ being the total angular momentum that has two components: the orbital angular momentum and the spins
\bea \label{eq:AMComponents}
\bm{J}_A^{\rm tot} = \bm{J}_A^{\rm orb} + \bm{J}_A^{\rm spin} = \bm{x}_A \times \bm{p}_A + \bm{S}_A. 
\eea
For the $\pi$-symmetric simulations we consider, the $\bm{J}^{\rm orb}_A$ are the same for the two black holes, but the $\bm{x}_A$ are opposite, so the orbital contribution from the two black holes cancel out in Eq.~\eqref{eq:CurrQuad}. The spin contribution, on the other hand, have opposite $\mathbb{P}\bm{S}_A$ for the two black holes ($\mathbb{P}$ is the projection operator into the orbital plane, which for the post-merger context will refer to the equatorial plane of the remnant black hole), and therefore does not need to vanish. Furthermore, the total orbital angular momentum $\sum_A \bm{J}_A^{\rm orb}$ and the total spin $\sum_A \bm{S}_A$ are both collinear with $\bm{J}_A^{\rm tot}$. 

Note that Eqs.~\eqref{eq:MassQuadrupole} and \eqref{eq:CurrQuad} are in the STF notation of Ref.~\cite{thorne80,PiraniBook,Sachs1961}, which has been summed over $m$. For our simulations, $\pi$-symmetry suppresses the $m=\pm 1$ modes, and even though there can be some small $m=0$ mode contribution in the waveforms, we are most interested in the $m=\pm 2$ modes. To approximate the $\bm{\mathcal{I}}$ and $\bm{\mathcal{S}}$ generating such modes in our simulations, we can use the quasi-Newtonian formula 
\bea \label{eq:MassDetail}
\bm{\mathcal{I}} = \frac{MR^2}{8}\bma 
 \cos(2\Omega \tilde{t}) + \frac{1}{3} &  \sin(2\Omega \tilde{t}) & 0 \\
 \sin(2\Omega \tilde{t}) & - \cos(2\Omega \tilde{t}) + \frac{1}{3} & 0 \\
0 & 0 & -\frac{2}{3} \ema
\eea
on a Cartesian coordinate system $(x,y,z)$ with $(x,y)$ spanning the orbital plane. The quantity $M$ is the total mass, $R$ is the separation between the black holes, and $\Omega =\sqrt{M/R^3}$ is the Newtonian orbital angular frequency. Note that we have replaced the time $t$ in a purely Newtonian expression by the retarded time $\tilde{t}$. 
For the current quadrupole, there are a few interesting configurations. First of all, when the spins are constant, anti-parallel to each other and in the orbital plane, we have 
\bea \label{eq:CurrConst}
\bm{\mathcal{S}} = \frac{R S}{2} \bma 
\frac{4}{3}\cos(\Omega \tilde{t}) & \sin(\Omega \tilde{t}) & 0 \\
\sin(\Omega \tilde{t}) & -\frac{2}{3}\cos(\Omega \tilde{t}) & 0 \\
0& 0 & -\frac{2}{3}\cos(\Omega \tilde{t})
\ema,
\eea 
where $S$ is the shared magnitude of the individual spins. Note that as the spins don't precess, there is only one $\Omega$ factor in Eq.~\eqref{eq:CurrConst} coming from the $\bm{x_A}$ term in Eq.~\eqref{eq:CurrQuad}, so the current quadrupole will evolve at the orbital frequency, instead of twice the orbital frequency like the mass quadrupole. If however, the spins precess also at frequency $\Omega$, we would then have the current quadrupole evolving at a frequency of $2\Omega$. For example, in the simple case where the spins are anti-parallel, locked to orthogonal directions to the line linking the black holes, and confined to the orbital plane, we have 
\bea \label{eq:CurrTransverse}
\bm{\mathcal{S}} = \frac{R S}{2} \bma 
-\sin(2\Omega\tilde{t}) & \cos(2\Omega\tilde{t}) & 0 \\
\cos(2\Omega\tilde{t}) & \sin(2\Omega\tilde{t}) & 0 \\
0& 0 & 0
\ema.
\eea
Another useful result is for the case when spin $\bm{S}_A$ is locked to the $-\bm{x}_A$ direction, where
\bea \label{eq:CurrLongitudinal}
\bm{\mathcal{S}} = -\frac{R S}{2} \bma
\cos(2\Omega\tilde{t}) + \frac{1}{3}  & \sin(2\Omega\tilde{t}) & 0 \\
\sin(2\Omega\tilde{t}) &-\cos(2\Omega\tilde{t})+ \frac{1}{3} & 0 \\
0& 0 & -\frac{2}{3}
\ema.
\eea
Finally, we note that if the two spins are aligned with each other, such as in the AA simulation (spins anti-aligned with the orbital angular momentum), then we suffer from the same effect that diminishes orbital contribution to $\bm{\mathcal{S}}$: the $\bm{S}_A$ are the same for the two black holes, while their $\bm{x}_A$ are opposite, so the overall current quadrupole vanishes. For the SK, SK- and SK$\bot$ simulations (spins initially in the orbital plane, see Table \ref{tb:superkickID}), the current quadrupole moment is non-vanishing during inspiral, and can be approximated by Eq.~\eqref{eq:CurrConst} during the early part of inspiral. Towards later stages of inspiral, the spin precession frequency increases and $\bm{\mathcal{S}}$ is somewhere between Eq.~\eqref{eq:CurrConst}, \eqref{eq:CurrTransverse} and \eqref{eq:CurrLongitudinal}. We now develop some tools for tracking how $\bm{\mathcal{S}}$ evolves in, e.g. the SK simulation, during the merger phase. 

The tendex and vortex fields corresponding to the current quadrupole $\bm{\mathcal{S}}$, in weak gravity, with a source region smaller than the gravitational wavelength,  
 are given in Ref.~\cite{Nichols:2011pu} as 
\bea 
\mathcal{B}_{ij} &=& \frac{2}{3}\left[-\left(\frac{\mathcal{S}_{pq}}{r} 
\right)_{,pqij} + \epsilon_{ipq} \left(\frac{\ddot{\mathcal{S}}_{pl}}{r} 
\right)_{,qk}\epsilon_{jlk} \right. \notag\\
&&\left. + 2\left(\frac{\ddot{\mathcal{S}}_{p(i}}{r} \right)_{,j)p}
- \left(\frac{\ddddot{\mathcal{S}}_{ij}}{r} \right)  \right],
\label{eq:CurrQuadVort}
\eea
\bea
\mathcal{E}_{ij} = \frac{4}{3} \epsilon_{pq(i}\left[-\left(\frac{\dot{
\mathcal{S}}_{pk}}{r}\right)_{,j)kq}+\left(\frac{\dddot{\mathcal{S}}_{j)p}}{r}
\right)_{,q}\right],
\label{eq:CurrQuadTend}
\eea
where repeated indices are summed over, and the overdot denotes time derivatives. Roughly, each time derivative introduces an $\Omega$ factor, while each spatial derivative can introduce either an $1/r$ factor when operating on explicit $r$'s in Eq.~\eqref{eq:CurrQuadVort} and \eqref{eq:CurrQuadTend}, or an $\Omega$ factor through the retarded time. In the near zone, where $r < \lambdabar$ ($\lambdabar$ is the reduced wavelength), the spatial derivatives that generate $1/r$ factors are favorable, so terms with more spatial derivatives are more dominant, and the strength of the tendex and vortex fields are determined by the first terms in Eqs.~\eqref{eq:CurrQuadVort} and \eqref{eq:CurrQuadTend}. The ratio of the strength between them is proportional to $\lambdabar/r$. When $r > \lambdabar$ (in the wave zone), it is preferable for spatial derivatives to introduce an $\Omega$ factor instead, and all the terms in the sums contribute equally. The result in the case of Eq.~\eqref{eq:CurrQuadVort} is essentially the same as a transverse-traceless projection operator acting on the four time derivative term multiplied by $-2$ \cite{AaronPrivate}. In this case, the $\bm{\mathcal{B}}$ and $\bm{\mathcal{E}}$ fields are of the same strength, as one would expect from them being sustained by mutual induction in a gravitational wave \cite{Nichols:2011pu}. 

The $\bm{\mathcal{E}}$ and $\bm{\mathcal{B}}$ fields generated by the mass quadrupole $\bm{\mathcal{I}}$  are the mirror image, and are given by 
\bea
\label{eq:MassQuadVort}
\mathcal{B}_{ij} = \epsilon_{pq(i}\left[\left(\frac{\dot{\mathcal{I}}_{pk}}{r}
\right)_{,j)kq}-\left(\frac{\dddot{\mathcal{I}}_{j)p}}{r}\right)_{,q}\right],
\eea
\bea
\label{eq:MassQuadTend}
\mathcal{E}_{ij} &=& \frac{1}{2}\left[-\left(\frac{\mathcal{I}_{pq}}{r} 
\right)_{,pqij} + \epsilon_{ipq} \left(\frac{\ddot{\mathcal{I}}_{pl}}{r} 
\right)_{,qk}\epsilon_{jlk} \right. \notag\\
&&\left. + 2\left(\frac{\ddot{\mathcal{I}}_{p(i}}{r} \right)_{,j)p}
- \left(\frac{ \ddddot{\mathcal{I}}_{ij}}{r} \right)  \right].
\eea

\begin{figure}[t,b]
\begin{overpic}[width=0.49\columnwidth]{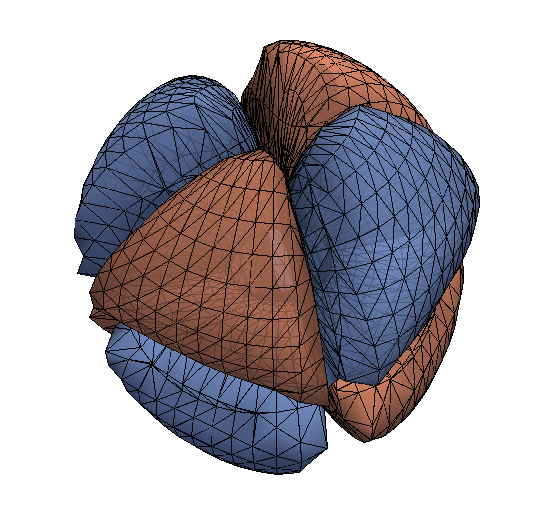}
\put(85,5){(a)}
\end{overpic}
\begin{overpic}[width=0.49\columnwidth]{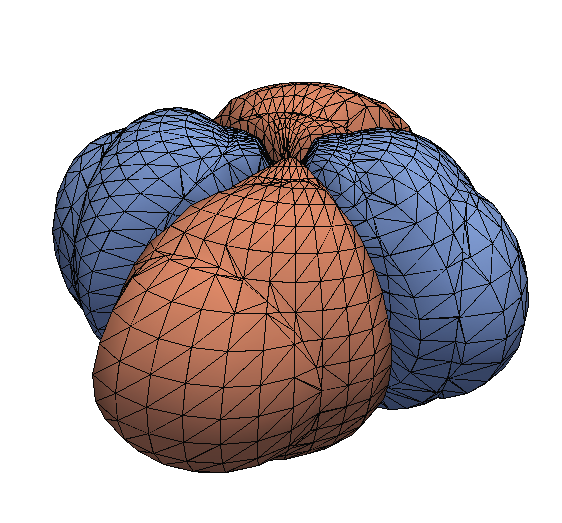}
\put(85,5){(b)}
\end{overpic}
\caption{
(a): Two Coulomb branch vorticity contours from the mass quadrupole according to Eq.~\eqref{eq:MassDetail}. The red and blue contours correspond to a pair of opposite vorticity values, with the red being $+$ve.
(b): Coulomb branch vorticity contours from the current quadrupole given by Eq.~\eqref{eq:CurrConst}.
}
\label{fig:VortexMassAndCurrent}
\end{figure}

The symmetry between Eqs.~\eqref{eq:CurrQuadVort}, \eqref{eq:CurrQuadTend}
and Eqs.~\eqref{eq:MassQuadVort}, \eqref{eq:MassQuadTend} allows for the definition of a complex quadrupole moment
\bea \label{eq:QuadComplex}
\mathcal{M}_{ij}= \frac{4}{3} \frac{\mathcal{S}_{ij}}{r}-i \frac{\mathcal{I}_{ij}}{r},
\eea
and then the tendex and vortex fields are given by the unified expression
\bea
\mathcal{E}_{ij} + i \mathcal{B}_{ij} &=& \epsilon_{pq(i}
\left[ -\dot{\mathcal{M}}_{pk,j)kq} +
\dddot{\mathcal{M}}_{j)p,q}
\right] \notag \\
&& + \frac{i}{2}\left[
-\mathcal{M}_{pq,pqij}
+\epsilon_{ipq}\ddot{\mathcal{M}}_{pm,qn} \epsilon_{jmn} \notag \right. \\
&& \left. + 2\ddot{\mathcal{M}}_{p(i,j)p}
+\ddddot{\mathcal{M}}_{ij}
\right].
\eea

We now turn to examine the symmetry properties of the $\bm{\mathcal{B}}$ field generated by Eqs.~\eqref{eq:CurrQuadVort} and \eqref{eq:MassQuadVort}. Following Ref.~\cite{Nichols:2012jn}, we define a positive/negative (abbreviated to $+$ve/$-$ve below) parity tensor field to be one that does not/does change sign under a reflection against the origin. 
Note that the parity operation we consider applies to the field location coordinates (e.g. $r$ in Eq.~\ref{eq:CurrQuadVort}), and not the source (black hole) locations or motions (e.g. $\bm{x}_A$ or $\bm{S}_A$ in Eq.~\ref{eq:AMComponents}), which can be seen formally as existing in a separate internal vector space.
This is akin to applying the parity transformation to only the $x$ coordinate of a Green function $G(x,x')$ while leaving $x'$ unaffected. 
Therefore, even though quantities like $\bm{x}_A$ are polar-vectors in that internal space, they, together with axial-vectors in the internal space, behave as axial-vectors under our parity transformation. Subsequently both $\bm{\mathcal{S}}$ and $\bm{\mathcal{I}}$ have $+$ve parity, as do
$\bm{\mathcal{B}}$ in Eq.~\eqref{eq:CurrQuadVort} and $\bm{\mathcal{E}}$ in Eq.~\eqref{eq:MassQuadTend} that have even numbers of derivatives, while $\bm{\mathcal{E}}$ in Eq.~\eqref{eq:CurrQuadTend} and $\bm{\mathcal{B}}$ in Eq.~\eqref{eq:MassQuadVort} take on $-$ve parity.

We show in Fig.~\ref{fig:VortexMassAndCurrent}, the Coulomb branch vorticity contours for $\bm{\mathcal{B}}$ as given by Eqs.~\eqref{eq:CurrQuadVort} and \eqref{eq:MassQuadVort}.
In addition to parity, our $\bm{\mathcal{B}}$ fields are $\pi$-symmetric by construction. So by combining a $\pi$-rotation with a parity transformation, we arrive at reflection anti-symmetry/symmetry against the orbital plane, for the mass/current quadrupole generated vorticity.
Furthermore, for the $m=\pm2$ modes we are considering, there is a $\pi/2$-rotation antisymmetry for both $\bm{\mathcal{I}}$ and $\bm{\mathcal{S}}$ generated vorticity, as evident from Fig.~\ref{fig:VortexMassAndCurrent}. We will call the combination of a parity transformation with a $\pi/2$-rotation the ``skew-reflection'', and then the mass/current quadrupole generated vorticity is skew-reflection symmetric/anti-symmetric.
Now consider an axisymmetric dipolar vorticity which also has a $-$ve parity, such as that generated by the orbital angular momentum or the spin of the post-merger final remnant black hole. It would be reflection anti-symmetric, as well as skew-reflection anti-symmetric. Therefore, a combination of the mass quadrupolar and the dipolar vorticities would have a definitive reflection anti-symmetry, but has no definitive skew-reflection symmetry property. Subsequently, the contours of opposite vorticities will be aligned with each other (in terms of rotation against the $\bm{J}^{\rm tot}$ axis) across the orbital plane. On the other hand, when we combine the current quadrupolar vorticity with the dipolar vorticity, we will have definite skew-reflection anti-symmetry, but no definitive reflection symmetry property. Therefore, the contours of opposite vorticities will be misaligned by $\pi/2$ instead.

\begin{figure}[!t]
\begin{overpic}[width=0.45\columnwidth]{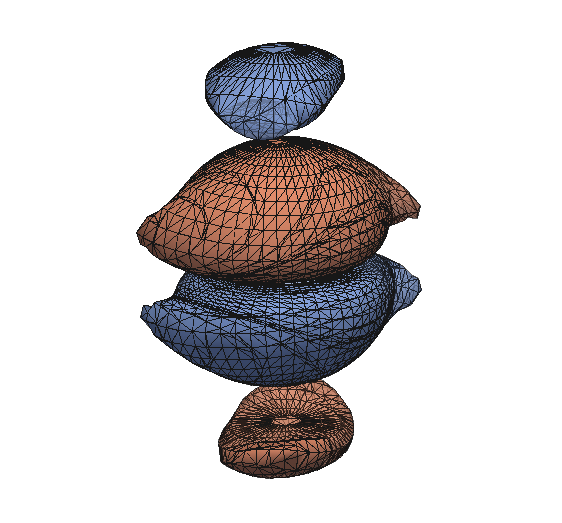}
\put(85,5){(a)}
\end{overpic}
\begin{overpic}[width=0.45\columnwidth]{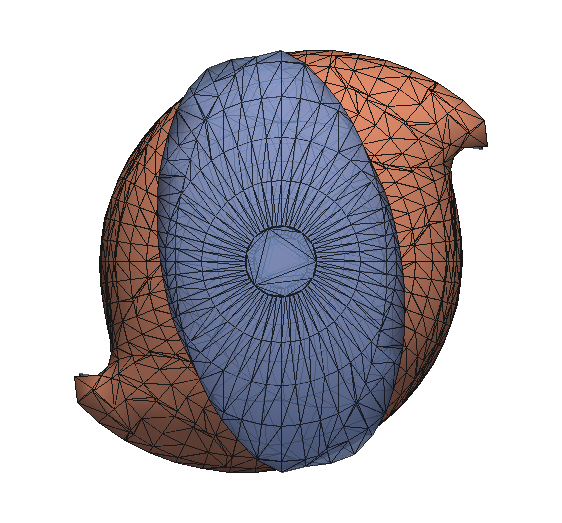}
\put(85,5){(b)}
\end{overpic}
\begin{overpic}[width=0.45\columnwidth]{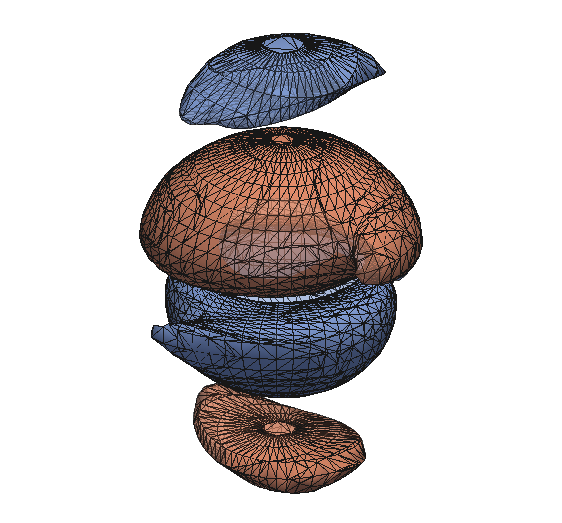}
\put(85,5){(c)}
\end{overpic}
\begin{overpic}[width=0.45\columnwidth]{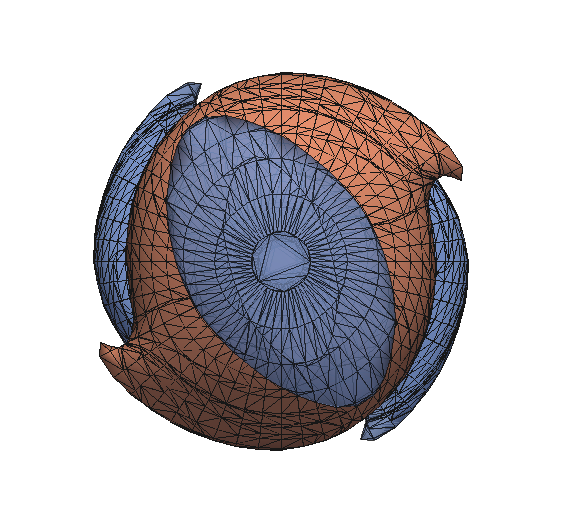}
\put(85,5){(d)}
\end{overpic}
\begin{overpic}[width=0.45\columnwidth]{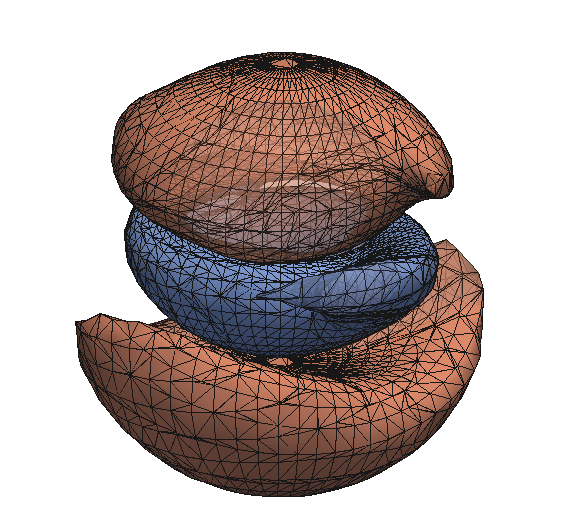}
\put(85,5){(e)}
\end{overpic}
\begin{overpic}[width=0.45\columnwidth]{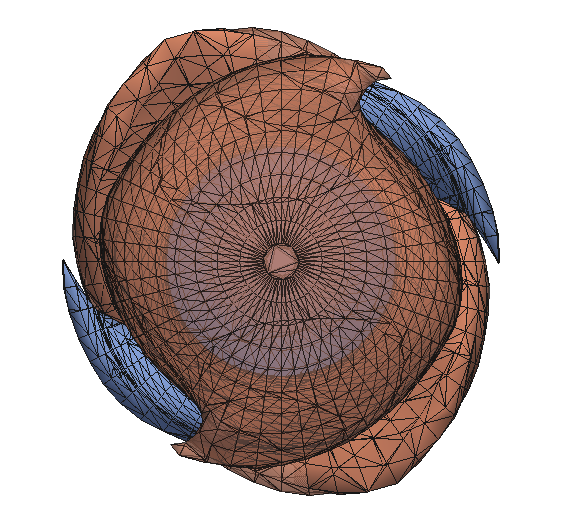}
\put(85,5){(f)}
\end{overpic}
\caption{The analytically constructed contours of opposite vorticity.
(a)-(b): Two contours of opposite vorticity from the mass quadrupole (Eq.~\ref{eq:MassDetail}) plus a current dipole. 
The contours, in particular the spiraling arms, are aligned with each other across the orbital plane. 
(c)-(d): Similar contours from the current quadrupole (Eq.~\ref{eq:CurrTransverse}) plus the dipole.  
The contours are misaligned by $\pi/2$, and obey skew-reflection antisymmetry.
(e)-(f): Both current and mass quadrupoles are included in addition to the dipole.
The contours are misaligned by an angle between $0$ and $\pi/2$, breaking both reflection and skew-reflection antisymmetry.  
Left column: side views of the contours (not at the same angle).
Right column: top views (looking down $\bm{J}^{\rm tot}$) of the contours. 
}
\label{fig:Analytical}
\end{figure}

This conclusion is demonstrated graphically in the top two rows of Fig.~\ref{fig:Analytical}, where contours of opposite vorticity are represented in red and blue. When constructing this figure, the dipole contribution is approximated as the vorticity field of a Kerr black hole in the Kerr-Schild coordinates.
When we combine the contributions from both quadrupoles as well as the dipole, the red and blue spiraling arms subtend a misalignment angle between $0$ and $\pi/2$ (possessing neither definitive reflection symmetry nor definitive skew-reflection symmetry), as shown in the third row of Fig.~\ref{fig:Analytical}. Note that although we have been utilizing weak gravity expressions in this section to construct examples, the qualitative symmetry considerations should remain valid in strong gravity, where this misalignment angle can serve as a convenient measure of the relative strength between the two types of quadrupoles.
Aside from a non-vanishing misalignment angle, another indicator for the existence of a current quadrupole contribution is that the contours can slice through the orbital plane (see Fig.~\ref{fig:VortexMassAndCurrent} (b) and Fig.~\ref{fig:Analytical} (e)), which is allowed by the skew-reflection anti-symmetry. On the other hand, reflection antisymmetry prevents the contours with non-vanishing vorticity from intersecting the orbital plane (see Fig.~\ref{fig:VortexMassAndCurrent} (a) and Fig.~\ref{fig:Analytical} (a)).
Finally, note that when both types of quadrupoles are present, the red and blue contours do not need to share the same size and/or shape (see Fig.~\ref{fig:VortexMassAndCurrent} (e) and (f)), with the difference between them dependent on the relative strength of these quadrupoles, as well as their relative phase.

\begin{figure}[!t]
\begin{overpic}[width=0.45\columnwidth]{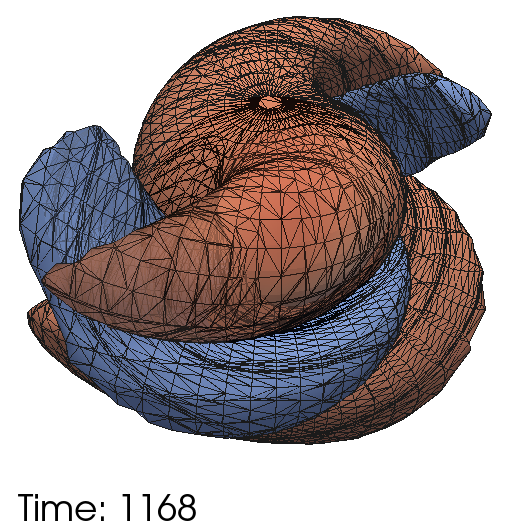}
\put(85,5){(a)}
\end{overpic}
\begin{overpic}[width=0.45\columnwidth]{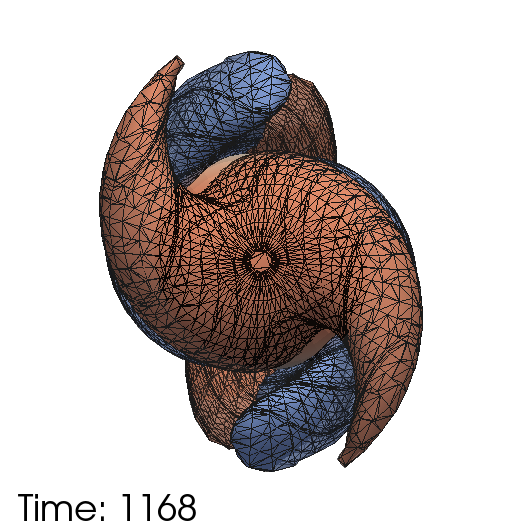}
\put(85,5){(b)}
\end{overpic}
\begin{overpic}[width=0.45\columnwidth]{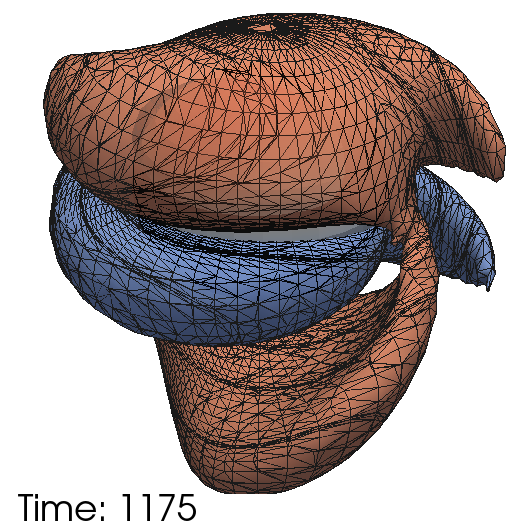}
\put(85,5){(c)}
\end{overpic}
\begin{overpic}[width=0.45\columnwidth]{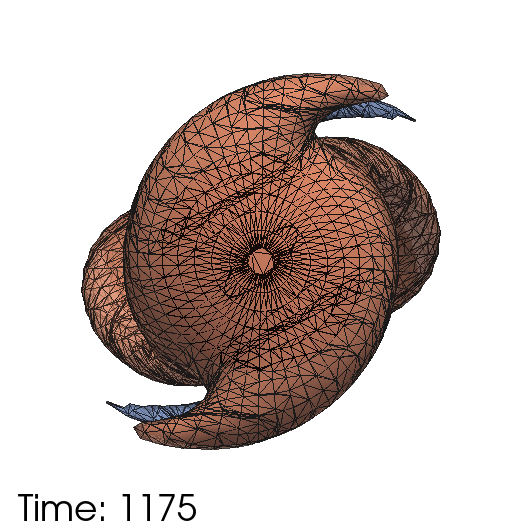}
\put(85,5){(d)}
\end{overpic}
\begin{overpic}[width=0.45\columnwidth]{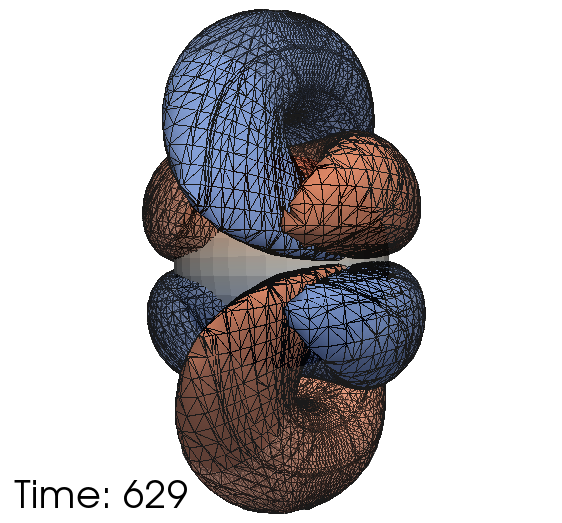}
\put(85,5){(e)}
\end{overpic}
\begin{overpic}[width=0.45\columnwidth]{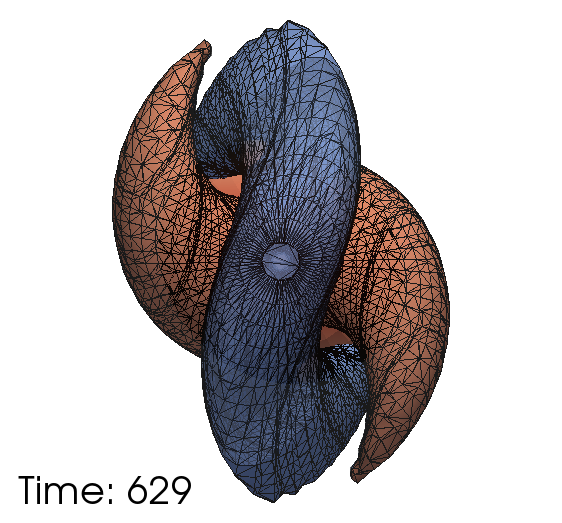}
\put(85,5){(f)}
\end{overpic}
\caption{The numerically observed contours of opposite vorticity in the simulations. (a)-(d): The time evolution of the Coulomb branch vorticity during the merger phase for the SK simulation. The blue and red surfaces are contours of the same absolute vorticity value but of the opposite signs. The rows correspond to different times.  Over time, the red and blue contours become more aligned with each other in their orientation.
(e)-(f): The Coulomb branch vorticity contours for the AA simulation, at the same time delay from merger as (a) and (b). The lack of current quadrupole moment ensures the alignment of the red and blue contours. 
Left column: side views of the contours (not at the same angle).
Right column: top views (looking down $\bm{J}^{\rm tot}$) of the contours.
}
\label{fig:EvolutionOfContourPostMerger}
\end{figure}

We now plot the time evolution (during the merger phase) of the opposite vorticity contours for the SK simulation in the top two rows of Fig.~\ref{fig:EvolutionOfContourPostMerger}, and observe an increase of alignment over time, suggesting a decreasing current quadrupole moment. We also note that the blue spiraling arms in Fig.~\ref{fig:EvolutionOfContourPostMerger} (a) slice through the orbital plane, while their counterparts in Fig.~\ref{fig:EvolutionOfContourPostMerger} (c) do not behave in the same way. Furthermore, the blue arms are larger in spatial extend initially, but reduce to be of similar sizes as the red arms later. These observations are also consistent with a declining current quadrupole contribution. 
For comparison, we also plot in the bottom row of Fig.~\ref{fig:EvolutionOfContourPostMerger}, the contours for the AA simulation at the same time delay from merger as in Fig.~\ref{fig:EvolutionOfContourPostMerger} (a) and (b), where as the current quadrupole moment vanishes according to Eq.~\eqref{eq:AMComponents}, the contours are already exactly aligned.

\section{Avenue for the exit of the current quadrupole \label{sec:ExitStrategy}}
If we hold spin magnitude $S$ constant, and keep spin directions tangential to the orbital plane in a $\pi$-symmetry configuration, then a comparison between Eq.~\eqref{eq:MassDetail} and Eqs.~\eqref{eq:CurrConst}-\eqref{eq:CurrLongitudinal} suggests that as $R$ decreases, the relative strength of $\bm{\mathcal{S}}$ as compared to $\bm{\mathcal{I}}$ should increase. Using the values for $M$ and $S$ in Table \ref{tb:superkickID}, we find that $\bm{\mathcal{S}}$ and $\bm{\mathcal{I}}$ should become similar in magnitude when $R \approx 2$, or around merger time. 
Furthermore, if $r < \lambdabar$ for the $r$ range we are interested in (e.g. the range plotted in Fig.~\ref{fig:EvolutionOfContourPostMerger}), then the first terms dominate in Eqs.~\eqref{eq:CurrQuadVort} and \eqref{eq:MassQuadVort}, and the ratio of $\bm{\mathcal{B}}$ field strength as generated by $\bm{\mathcal{S}}$ to that generated by $\bm{\mathcal{I}}$ is multiplied by a factor of $\lambdabar /r >1$ on top of the strength ratio between $\bm{\mathcal{S}}$ and $\bm{\mathcal{I}}$. On the other hand, when $r > \lambdabar$, all the terms in Eqs.~\eqref{eq:CurrQuadVort} and \eqref{eq:MassQuadVort} contribute, so the additional factor is around $1$ because all the terms introduce an $\Omega^4$ factor. In other words, the current quadrupole is either more effective or equally effective at generating the vortex field as compared to the mass quadrupole. This conclusion should remain true in a strong gravitational field, as $\bm{\mathcal{B}}$ can be seen as the primary field generated by $\bm{\mathcal{S}}$, while only a secondary field for $\bm{\mathcal{I}}$ that is induced by the time variation of $\bm{\mathcal{E}}$ (see Sec.~VI C and Sec.~VI D 1 of Ref.~\cite{Nichols:2011pu} for further discussions and examples). 
Consequently, as the absolute magnitude of $\bm{\mathcal{S}}$ catches up to or even overtakes that of $\bm{\mathcal{I}}$ during the merger phase, we should not see its influence in vorticity decline in the fashion of Fig.~\ref{fig:EvolutionOfContourPostMerger}. Therefore, the current quadrupole must somehow diminish during the merger phase.  
It would vanish if the equal and opposite spins of the two individual black holes simply annihilate each other, but this does not explain why spins annihilate faster than the two masses merge (i.e. why the current quadrupole declines faster than the mass quadrupole). In other words, we need the current quadrupole to reduce faster than the signature of the individual black holes disappears. 

Such a scenario is possible if the individual spins experience a re-orientation into configurations that produce near-vanishing $\bm{\mathcal{S}}$, even as the spins themselves are still non-vanishing. This can be achieved if the spins move to become nearly aligned with each other, which according to Eq.~\eqref{eq:AMComponents} would result in a nearly vanishing $\bm{\mathcal{S}}$.

Above, we have used the term ``individual spins'' in a generalized sense. Even before merger, the spins of the individual black holes are really reflections of the spacetime dynamics outside of these holes, as the characteristic modes for the Einstein equation are strictly outgoing at the apparent horizons (which are inside the event horizons). The merger would not instantaneously remove the near-zone dynamics that were underlying the individual spins, so one may regard the continuation of such dynamics as a kind of remnant spin (the word ``remnant'' will be omitted frequently for brevity). In the tendex and vortex language, one may say that these remnant spins are the vorticity and tendicity of the spacetime that were associated with the spins before merger, but would not instantly dissipate upon the formation of the common apparent horizon. Similarly, the tendicity and the vorticity of the spacetime that were associated with the orbital motion of the individual black holes before merger continue to evolve post-merger, and provide a remnant orbital motion. We note that the remnant spins considered here belong to the defunct individual black holes, and are not the spin of the remnant black hole.

Due to the lack of analytical descriptions for highly dynamic regimes, we will rely on the available perturbative expressions to aid our qualitative arguments in the rest of this section. Although we are likely pushing these expressions beyond their reasonable range of validity, we will only be interested in the qualitative features of the spacetime they expose, and not their quantitative accuracy. 

In order to achieve alignment, it is required that the spins be lifted out of the orbital plane, either through spin-orbit coupling or spin-spin coupling, because the $\pi$-symmetry forbids the spins from being aligned when they are confined to the orbital plane.
The spin-orbit coupling is given by the leading order PN expression 
\cite{Barker:1975ae,1979GReGr..11..149B,1993PhRvD..47.4183K,Kidder:1995zr}
\bea \label{eq:SpinOrbit}
\bm{\dot{S}}_1 = \frac{1}{R^3}\left(2+\frac{3}{2}\frac{m_2}{m_1}\right)\left(\bm{L}_N \times
\bm{S_1} \right),
\eea
in the center-of-mass frame, where $\bm{L}_N$ is the orbital angular momentum at the Newtonian order
\bea
\bm{L}_N = \mu \bm{R} \times \dot{\bm{R}}
\eea
with $\mu$ being the reduced mass and $\bm{R} =\bm{x}_1 -\bm{x}_2$, and also $R$ being the magnitude of $\bm{R}$. For the post-merger context, we will instead take $\bm{L}_N$ to be $\bm{J}^{\rm tot}$ minus the remnant individual spins. We note that the $\bm{\dot{S}}_1$ in Eq.~\eqref{eq:SpinOrbit} cannot point out of the orbital plane and will only generate spin precession within it. 

The spin-spin coupling, on the other hand, can create a torque pointing out of the orbital plane. 
The leading order expression for spin-spin coupling is given by \cite{Barker:1975ae,1979GReGr..11..149B,1993PhRvD..47.4183K,Kidder:1995zr}
\bea \label{eq:SpinSpin1PN}
\dot{\bm{S}}_1 = -\frac{1}{R^3}\left[\bm{S}_2 - 3(\bm{n}\cdot \bm{S}_2)\bm{n} \right] \times \bm{S}_1,
\eea
in the center-of-mass frame, and $\bm{n}$ is defined as $\bm{R}/R$. 
For the SK configuration, $\bm{S}_2 \times \bm{S}_1 =0$ when the spins are in the orbital plane, as they must be anti-parallel by $\pi$-symmetry, but $\bm{S}_2$ is not required to be transverse (especially when the spins are not locked to the orbital motion), so $\bm{n}\cdot \bm{S}_2 \neq 0$ and there is a $\bm{\dot{S}}$ in the direction orthogonal to the orbital plane. Normally, this direction is not constant over an orbital cycle for a pair of spins not locked to the orbital motion, so the spin-spin coupling effect does not accumulate significantly during early inspiral, but as the merger phase will not take up a whole cycle, we need not worry about cancellations. 
On the other hand, the directions of $\bm{\dot{S}}_1$ and $\bm{\dot{S}}_2$ are the same, so as desired, the spins for our SK configuration either both move upwards (more aligned with $\bm{J}^{\rm tot}$) or both move downwards (more anti-aligned with $\bm{J}^{\rm tot}$), retaining the $\pi$-symmetry. 

According to Eq.~\eqref{eq:SpinSpin1PN}, the spins do not need to move towards spin-spin alignment. 
In fact, the $\pi$-symmetry enforces $\mathbb{P}\bm{S}_1 = -\mathbb{P}\bm{S}_2$, so unless $\mathbb{P}\bm{S}_1 = \mathbb{P}\bm{S}_2 = 0$, the spins would never be aligned. 
The situation would change however if we include the radiation reaction. Because gravitational waves drain dynamical energy, radiation reaction should push the spin orientations towards an energetically favorable equilibrium configuration, where the spin precession due to both spin-spin and spin-orbit coupling ceases. From Eq.~\ref{eq:SpinOrbit}, it is clear that the spin-orbit induced precession would shut off only when $\bm{S}_1$ and $\bm{S}_2$ are orthogonal to the orbital plane, or in other words are collinear with $\bm{J}^{\rm tot}$.
Furthermore, Eq.~\eqref{eq:SpinSpin1PN} shows that the spin-spin coupling generated spin precession also stops for such configurations. In addition, when the spins are either both aligned or both anti-aligned with $\bm{J}^{\rm tot}$ (henceforth referred to as spin-$\bm{J}^{\rm tot}$ alignment or anti-alignment), the current quadrupole will vanish.  

For this postulate to work, a necessary condition is that the spin-$\bm{J}^{\rm tot}$ alignment or anti-alignment configuration should correspond to a local minimum in energy. To this end, we note that the potential energy associated with the spin-spin interaction at leading order is given by Refs.~\cite{Barker:1975ae,1979GReGr..11..149B,1993PhRvD..47.4183K,Kidder:1995zr} as
\bea \label{eq:SSPotential}
U_{\text{SS}}=-\frac{1}{R^3}\left[\bm{S}_1 \cdot \bm{S}_2 -3\left(\bm{S}_1 \cdot \bm{n} \right) \left(\bm{S}_2 \cdot \bm{n} \right) \right],
\eea
while the potential for the spin-orbit coupling is \cite{Barker:1975ae,1979GReGr..11..149B,1993PhRvD..47.4183K,Kidder:1995zr}
\bea \label{eq:SOPotential}
U_{\text{SO}} = \frac{1}{R^3}\bm{L}_N \cdot \left(\left(2+\frac{3m_1}{2m_2}\right)\bm{S}_2 + \left(2+\frac{3m_2}{2m_1}\right)\bm{S}_1 \right).
\eea

The absolute minimum of $U_{\text{SS}}$ is achieved when $\bm{S}_1$ and $\bm{S}_2$ are anti-parallel and collinear with $\bm{n}$, as the second term in the square bracket of Eq.~\eqref{eq:SSPotential} favors anti-parallel orientations and dominates over the parallel-orientation favoring first term, due to its extra factor of $3$. 
This is also an equilibrium configuration for Eq.~\eqref{eq:SpinSpin1PN}, but does not lead to a small current quadrupole moment, as shown by Eq.~\eqref{eq:CurrLongitudinal}.

For the spin-$\bm{J}^{\rm tot}$ alignment or anti-alignment equilibrium configurations that we are interested in, we have $\phi_1 = \phi_2 = 0$ or $\pi$, and $\theta_1 = \theta_2 = \pi/2$, with $\theta_1$ being the angle $\bm{S}_1$ spans with $\bm{n}$ and $\phi_1$ the angle between $\bm{J}^{\rm tot}$ and the projection of $\bm{S}_1$ into the plane orthogonal to $\bm{n}$. The angles $\theta_2$ and $\phi_2$ are defined similarly for $\bm{S}_2$. It is easy to verify that all first derivatives of $U_{\text{SS}}$ against the angles vanish for these configurations, so they are indeed critical/equilibrium points. However, the eigenvalues of the Hessian are $\{0,2,3,-1\}|\bm{S}_1||\bm{S}_2|/R^3$ and not all positive, so they are not (local) minima of the potential energy.
When we add in the potential $U_{\text{SO}}$, which achieves its absolute minimum at the spin-$\bm{J}^{\rm tot}$ anti-alignment configuration, the eigenvalues of the Hessian (of $U_{\text{SO}} + U_{\text{SS}}$) for this configuration become
\bea
\frac{1}{4R^3} \left\{ 7{L}_N, 7{L}_N-1, 7{L}_N + 2, 7{L}_N + 3 \right\},
\eea
where we have taken $|\bm{S}_1|=|\bm{S}_2|=1/2$ and $\bm{L}_N = L_N \bm{\hat{z}}$ to simplify expressions. For our $M \approx 1$ simulations, and using the Newtonian expression for $\bm{L}_N$, spin-$\bm{J}^{\rm tot}$ anti-alignment configuration is a local minimum as long as $R >1/49$M. The spin-$\bm{J}^{\rm tot}$ alignment configuration, on the other hand, has eigenvalues
\bea
\frac{1}{4R^3} \left\{ -7{L}_N, -7{L}_N-1, -7{L}_N + 2, -7{L}_N + 3 \right\},
\eea
and is therefore not a local minimum unless ${L}_N$ is sufficiently negative.
When we add in the next-to-leading-order PN expressions for $U_{\text{SO}}$ \cite{Tagoshi-Ohashi-Owen:2001,Faye:2006gx,Porto:2010tr,Damour:024009,Levi:2010zu}, we acquire extra multiplicative factors onto $\bm{L}_N$ that can reverse the sign of the effective ${L}_N$ at small $r$, and make spin-$\bm{J}^{\rm tot}$ alignment configuration a local minimum (effective $\bm{L}_N$ in $U_{\text{SO}}$ is reversed, but $\bm{J}^{\rm tot}$ is not, so anti-alignment with the effective $\bm{L}_N$ now results in an alignment with $\bm{J}^{\rm tot}$).
The $U_{\text{SO}}$ that includes both leading and next-to-leading order PN contributions can be deduced from the spin precession equation (Eqs.~$61$-$64$ in Ref.~\cite{Porto:2010tr})
\bea \label{eq:PrecessionFull}
\dot{\bm{\bar{S}}}_1 = \bm{H}_1 \times \bm{\bar{S}}_1
\eea
in a general frame, where we can regard $\bm{H}_1 R^3 / (2+3m_2/2m_1)$ as an effective $\bm{L}_N$ for $\bm{\bar{S}}_1$, and $\bm{S}_1$'s contribution to $U_{\text{SO}}$ is $\bm{H}_1 \cdot \bm{\bar{S}}_1$. The quantities appearing in Eq.~\eqref{eq:PrecessionFull} are
\bea
\bm{\bar{S}}_1 &=& \left( 1- \frac{\bm{\bar{v}}^2_1}{2} - \frac{\bm{\bar{v}}^4_1}{8} \right) \bm{S}_1 \notag \\
&&+ \frac{1}{2}\bm{\bar{v}}_1(\bm{\bar{v}}_1\cdot \bm{S}_1)\left( 1+ \frac{1}{4}\bm{\bar{v}}^2_1\right) \label{eq:EffSpin} \\ 
\bm{\bar{v}}_1 &=& \frac{\left(1+m_2/R\right) \bm{v}_1 - 2(m_2/R) \bm{v}_2}{1-m_2/R}
\eea
and more importantly
\begin{widetext}
\bea 
\bm{H}_1 &=& \frac{m_2}{R^3}\left[\left( \frac{3}{2} + \frac{1}{8}  \bm{v}^2_1 +  \bm{v}^2_2 -  \bm{v}_1 \cdot  \bm{v}_2 - \frac{9}{4}(\bm{v}_2\cdot \bm{n})^2 + \frac{1}{2R}(m_1 -m_2) \right) \bm{L}_1 \right. \notag \\
&& \left. + \left( -2 - 2 \bm{v}^2_2 + 2\bm{v}_1 \cdot  \bm{v}_2 + 3(\bm{v}_2\cdot \bm{n})^2 + \frac{1}{2R}(2m_1 +3 m_2)  \right) \bm{L}_2
- \frac{1}{2} (\bm{v}_2 \cdot \bm{R}) \bm{v}_1 \times \bm{v}_2 \right] \label{eq:NLOUSS}
\eea
\end{widetext}
with $\bm{L}_1 = \bm{R} \times \bm{v}_1$ and $\bm{L}_2 = \bm{R} \times \bm{v}_2$. Note $\bm{L}_2$ is in the opposite direction to particle $2$'s orbital angular momentum, so the 
$(1/2R)(2m_1 +3 m_2)$ term in Eq.~\eqref{eq:NLOUSS} could reverse the direction of the effective $\bm{L}_N$ when $R$ is small. An interesting complication is 
the fact that the equilibrium direction as determined by $\bm{H}_1$ is for $\bm{\bar{S}}_1$ that has a directional difference with $\bm{S}_1$ from the term proportional to $\bm{\bar{v}}_1$ in Eq.~\eqref{eq:EffSpin}, and that we can also have directional adjustments in $\bm{H}_1$ to make it deviate from the $\pm \bm{J}^{\rm tot}$ directions.  
Such frame-dependent adjustments may prevent a perfect alignment or anti-alignment of the spins with $\bm{J}^{\rm tot}$, and so the current quadrupole will not vanish completely.  
We will speculate on the significance of this complication in Sec.~\ref{sec:Discussion}. 

Throwing away the crutch of perturbative expressions, we really only need the qualitative statements, that the spin-$\bm{J}^{\rm tot}$ near-alignment or near-antialignment configurations are energetically favorable for the spin-orbit coupling, and that mechanisms like the spin-spin coupling exist that can lift the spins out of the orbital plane, to remain true in the strong field regime. The perturbative expressions have hinted that it may be possible to meet these requirements, but strong field expressions are needed for quantitative assessments.

\begin{figure}[t,b]
\includegraphics[width=0.75\columnwidth]{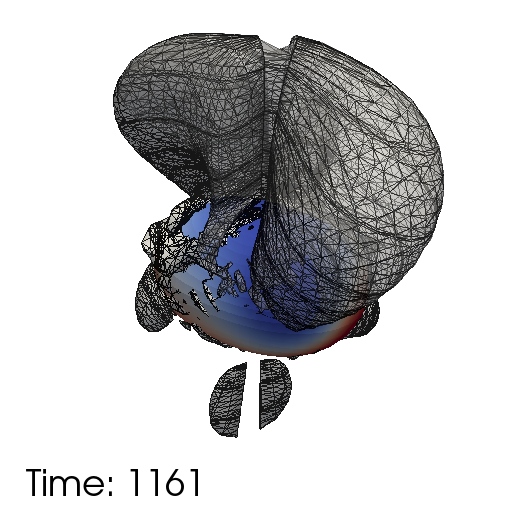}
\caption{A contour of $\Re(\Psi_4)$ for the SK simulation immediately after the merger, with $\Psi_4$ extracted on the quasi-Kinnersley tetrad. Also shown is the common apparent horizon colored by the horizon vorticity $\mathcal{B}_{NN}$. The $\Re(\Psi_4)$ contour connects to certain blue horizon vorticity patches.
These are the negative $\mathcal{B}_{NN}$ counterparts to the $C'$ patch of Fig.~\ref{fig:HorizonVorticity} (d) below, which we will show in Sec.~\ref{sec:ObRemSpin} to be a direct manifestation of the remnant spins. Note that there is some visualization complication near the polar directions, where we don't have any collocation points, so the contour shown there is an ill-constructed interpolation.
}
\label{fig:SourceOfRad}
\end{figure}

Another condition for our postulate to work is that the remnant spin dynamics should be efficient radiators during the merger phase, such that the spins experience a significant reaction. In contrast, when modeling early inspiral, the radiation reaction felt by the spins is usually neglected \cite{Kidder:1995zr,Apostolatos1994}. If we plot the distribution of $\Psi_4$ near the common apparent horizon, we should see a close association between the high intensity regions of $\Psi_4$ and entities that can be interpreted as representing the remnant spins. To verify this, we adopt the quasi-Kinnersley tetrad described in Refs.~\cite{Beetle2005,Nerozzi2005,Burko2006,Nerozzi:2005hz,Burko:2007ps,Zhang:2012ky}.
Our particular version of the tetrad follows Ref.~\cite{Zhang:2012ky} and suffers from some numerical noise because of the third derivative of the metric required for its construction. However, because we are now examining the region very close to the remnant black hole, the mixing of $\Psi_2$ into $\Psi_4$ under a simple simulation-coordinate based tetrad will overwhelm the interesting features. The quasi-Kinnersley tetrad avoids this problem by correctly identifying the gravitational wave propagation direction \cite{Zhang:2012ky}. Specifically, the tetrad bases correspond directly to the super-Poynting vector, so that the $\Psi_4$ extracted under this tetrad retains a simple relationship with the energy flux even in the near zone.   
In Fig.~\ref{fig:SourceOfRad}, we plot a large absolute value, and thus close to the radiating source, contour of $\Re(\Psi_4)$. It is clear that this contour attaches to certain blue horizon vorticity patches (contours of even higher $|\Re(\Psi_4)|$ are observed to be confined to regions closer to these patches), which we will now show (in Sec.~\ref{sec:ObRemSpin}) to be direct manifestations of the remnant spins.

\section{Observing the remnant spins \label{sec:ObRemSpin}}

\begin{figure*}[!ht]
\begin{overpic}[width=0.245\textwidth]{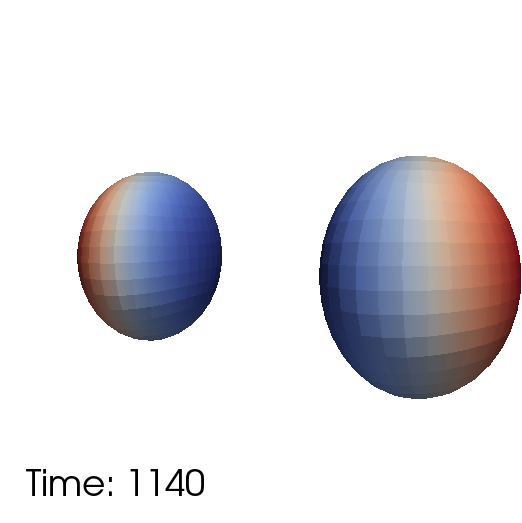}
\put(85,5){(a)}
\end{overpic}
\begin{overpic}[width=0.245\textwidth]{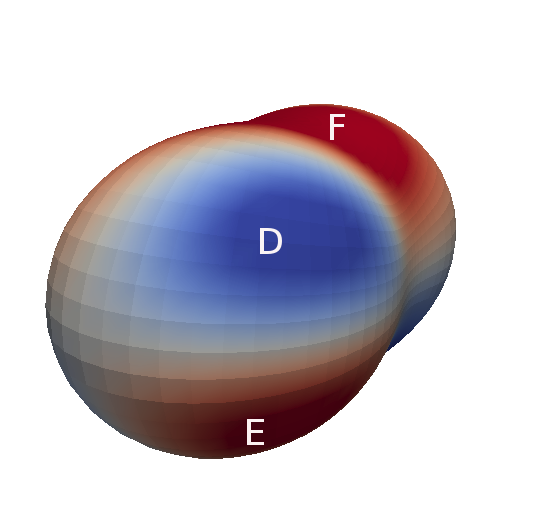}
\put(85,5){(b)}
\end{overpic}
\begin{overpic}[width=0.245\textwidth]{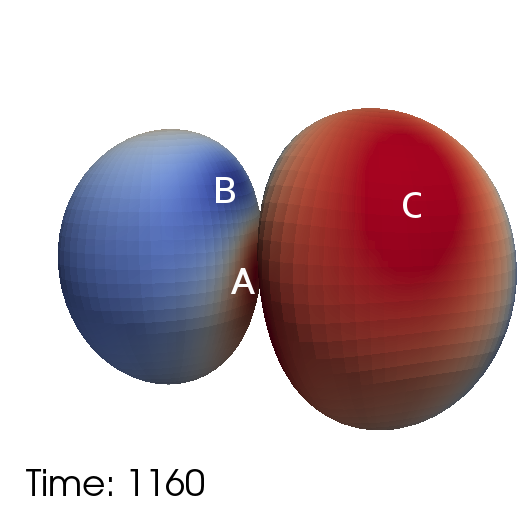}
\put(85,5){(c)}
\end{overpic}
\begin{overpic}[width=0.245\textwidth]{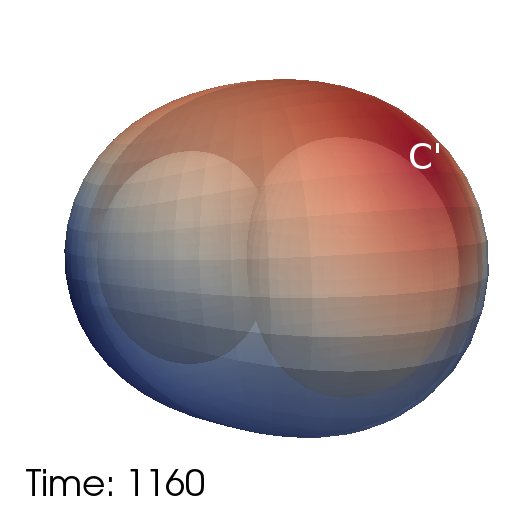}
\put(85,5){(d)}
\end{overpic}
\begin{overpic}[width=0.245\textwidth]{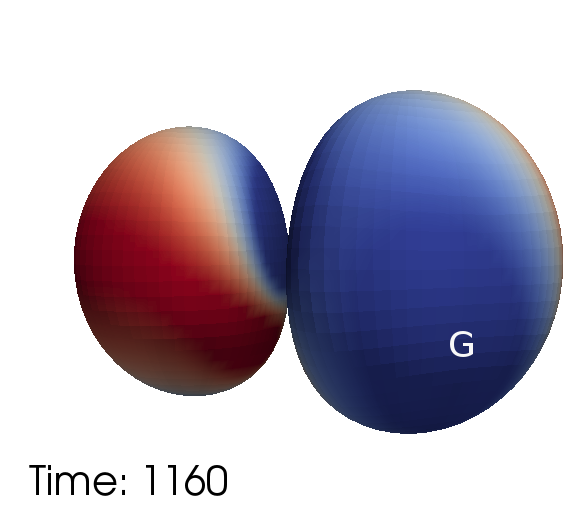}
\put(85,5){(e)}
\end{overpic}
\begin{overpic}[width=0.245\textwidth]{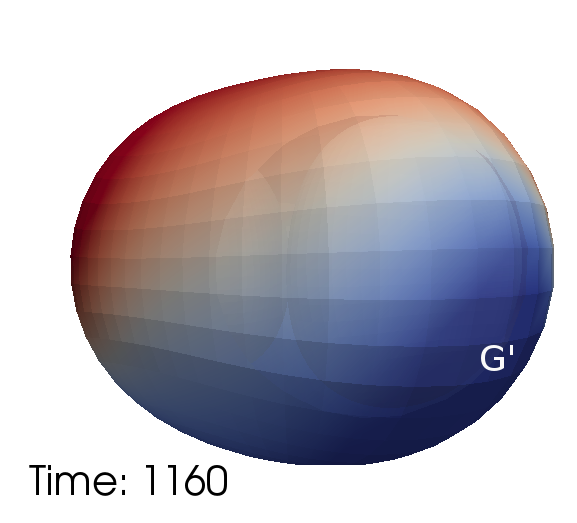}
\put(85,5){(f)}
\end{overpic}
\begin{overpic}[width=0.245\textwidth]{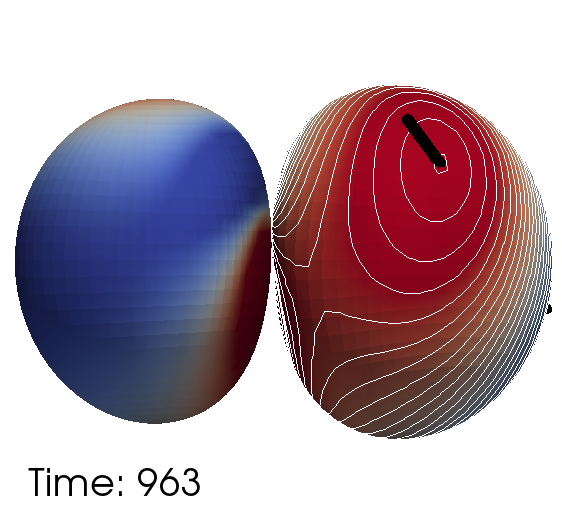}
\put(85,5){(g)}
\end{overpic}
\begin{overpic}[width=0.245\textwidth]{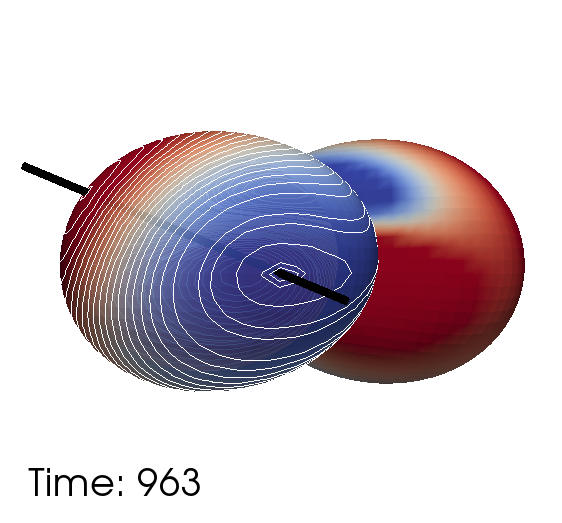}
\put(85,5){(h)}
\end{overpic}
\caption{
(a): The inspiral phase of the SK simulation, where the dipolar $\mathcal{B}_{NN}$ patterns on each individual horizon can be used to identify the spin direction. The red color corresponds to $+$ve $\mathcal{B}_{NN}$, and the spin of each black hole points from the blue patch towards the red patch. 
(b): The post-merger $\mathcal{B}_{NN}$ for the AA simulation. In addition to the dipolar patch $F$ due to the spin of the remnant black hole, the individual remnant spin patches $D$ and $E$ are also clearly visible, which shows that the spins remain in the anti-aligned orientation as expected. These spin patches retain this orientation throughout the merger phase. 
(c): Immediately before the merger in the SK simulation. In addition to the spin patch $C$, mass quadrupole generated patches $A$ and $B$ are also visible on the insides (the sides of the individual horizons that are facing each other).
(d): Same as (c) but includes the common apparent horizon as a semi-transparent surface. The patch $C'$ on the common horizon corresponds to the patch $C$ on the individual horizons.
(e)-(f): Similar to (c)-(d), but for the SK- simulation instead of the SK simulation. 
(g)-(h): Similar to (c), but for the SK$\bot$ simulation. (g) is the front view showing the $+$ve $\mathcal{B}_{NN}$ spin patch, while (h) is the side view showing the $-$ve $\mathcal{B}_{NN}$ spin patch. The white curves are the contours of $\mathcal{B}_{NN}$, and we have drawn a thick black line connecting the centers of the two spin patches. The orientation of this line is used as an approximate measure of the spin direction. 
}
\label{fig:HorizonVorticity}
\end{figure*}

\begin{figure*}[!ht]
\begin{overpic}[width=0.49\textwidth]{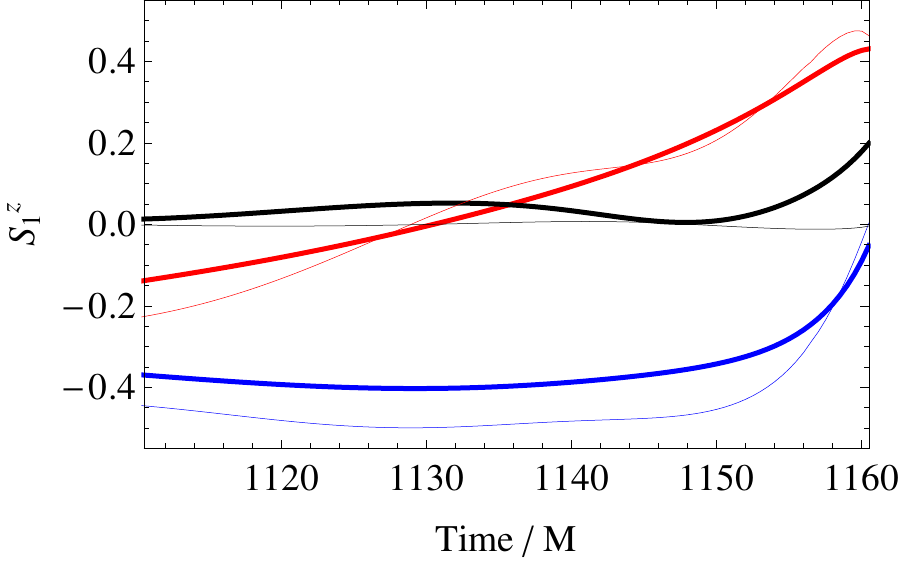}
\put(90,2){(a)}
\end{overpic}
\begin{overpic}[width=0.49\textwidth]{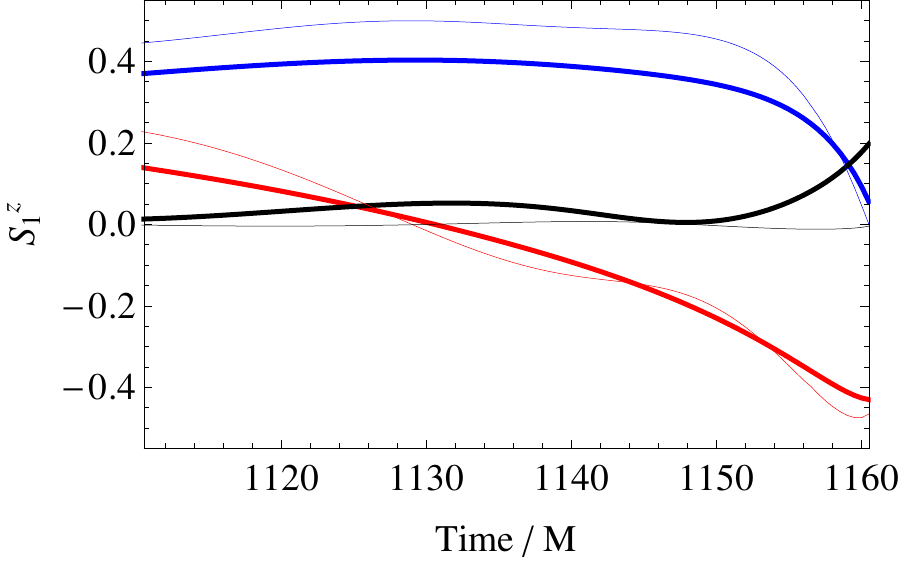}
\put(90,2){(b)}
\end{overpic}
\begin{overpic}[width=0.49\textwidth]{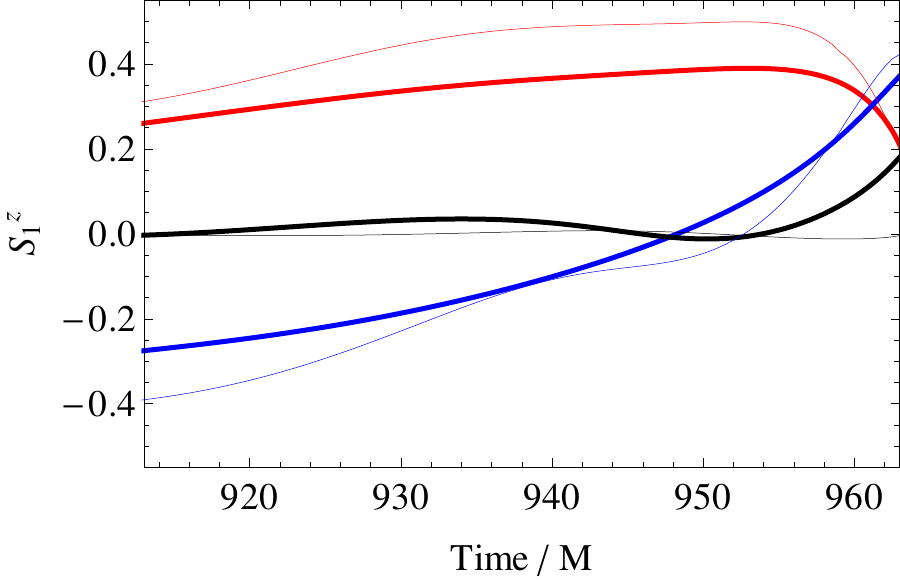}
\put(90,2){(c)}
\end{overpic}
\begin{overpic}[width=0.49\textwidth]{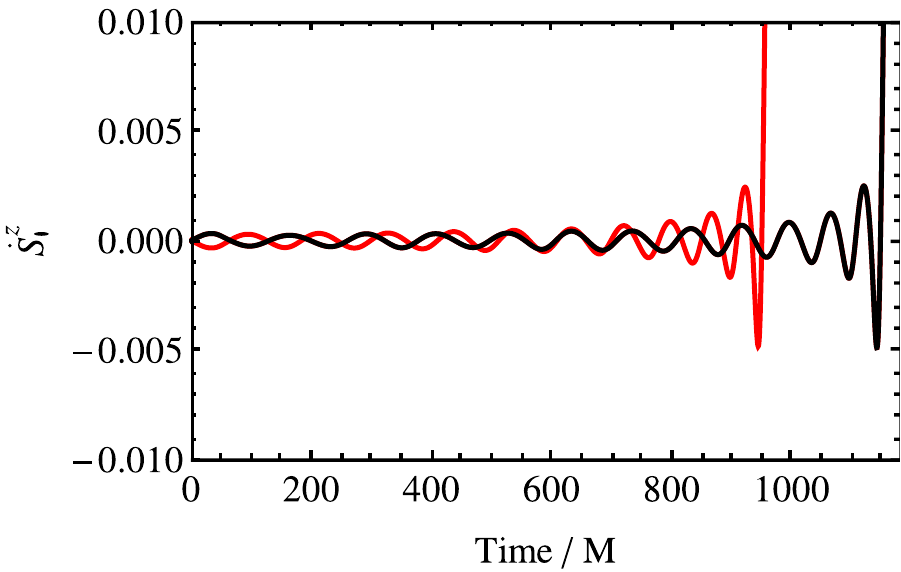}
\put(90,2){(d)}
\end{overpic}
\caption{The analytically predicted (thick lines) vs. the numerically computed (thin lines) $\bm{S}_1$ for the SK (a), SK- (b) and SK$\bot$ (c) simulations. The red, blue and black lines correspond to $S^x_1$, $S^y_1$ and $S^z_1$, respectively. We focus on the last $50$M before merger. Panel (d) shows the analytical predictions for $\dot{S}_1^z$ of the SK/SK- (both correspond to the black curve) and the SK$\bot$ (red curve) simulations.
}
\label{fig:AnalyticalSpin}
\end{figure*}

One possible way to visualize the (remnant) spins directly is through the horizon vorticity $\mathcal{B}_{NN}$. This quantity is 
closely related to the spin measurement expressed as an integral over the horizon \cite{BrownYork1993,Ashtekar2001,Ashtekar2003}, namely \cite{ZhangPRD2}
\bea \label{eq:HorSpin1}
S = \frac{1}{8\pi} \oint \mathcal{X} \zeta dA,
\eea
where $dA$ is the area element on the horizon, and for the event horizons, $\mathcal{X}$
is $-2\mathcal{B}_{NN}$ plus some spin coefficient corrections that vanish in a stationary limit (see Refs.~\cite{OwenEtAl:2011,ZhangPRD2}). We will use the apparent horizons in this paper, which coincide with the event horizons in the stationary limit \cite{CarterAHEH}, but are not teleological and therefore more widely utilized in numerical simulations. The quantity $\zeta$ in Eq.~\eqref{eq:HorSpin1} is determined by an eigenvalue problem on the horizon, and is essentially an $l=1$ spherical harmonic \cite{Owen:2009sb}. In other words, the spin is given by the dipole part of the horizon vorticity $\mathcal{B}_{NN}$. For example, during early inspiral, the $\mathcal{B}_{NN}$ pattern on each individual horizon is dominated by the spin of that black hole, forming a dipolar shape like those seen for the Kerr black holes in Ref.~\cite{ZhangPRD2}. This is shown in Fig.~\ref{fig:HorizonVorticity} (a), and can be used to identify the spin directions. 

Close to and past merger, aside from the overall $l=1$ harmonic-weighted integral in Eq.~\eqref{eq:HorSpin1}, there are many interesting finer details that can be seen from the $\mathcal{B}_{NN}$ plots for our various simulations shown in Fig.~\ref{fig:HorizonVorticity}. 
The panels (c)-(d), (e)-(f) and (g)-(h) depict the horizon vorticity patterns for the SK, SK- and SK$\bot$ simulations, respectively, while panel (b) shows the pattern for the AA simulation. The panels (b), (d) and (f) show the common apparent horizon, while the rest of the panels are for the two individual horizons. 
At the end of the inspiral stage, the horizon vorticity picks up a visible contribution from the mass quadrupole-induced $\bm{\mathcal{B}}$ field (e.g. patches $A$ and $B$ in Fig.~\ref{fig:HorizonVorticity} (c) for the SK simulation), while the spin contributions are also present (e.g. patch $C$ and a blue patch at the back that is blocked from view in Fig.~\ref{fig:HorizonVorticity} (c)). 
The spin patches like $C$ are smooth continuations of the dipolar patches we see during early inspiral in Fig.~\ref{fig:HorizonVorticity} (a), while mass quadrupole patches $A$ and $B$ only appear shortly before merger, but grow quickly to have larger $|\mathcal{B}_{NN}|$ than the spin patches. 

Post-merger, the horizon vorticity for the SK simulation is shown in Fig.~\ref{fig:HorizonVorticity} (d), where in addition to a dipolar contribution from the spin of the remnant black hole (similar to the $F$ patch in Fig.~\ref{fig:HorizonVorticity} (b) for the AA simulation), there are also visible patches that can be interpreted as the continuation of the pre-merger spin patches. For example, the region $C'$ in Fig.~\ref{fig:HorizonVorticity} (d) corresponds to patch $C$ in Fig.~\ref{fig:HorizonVorticity} (c), while the patches $D$ and $E$ in Fig.~\ref{fig:HorizonVorticity} (b) correspond to the continuation of the pre-merger individual spins in the AA simulation that were anti-aligned with $\bm{J}^{\rm tot}$. Such finer details in $\mathcal{B}_{NN}$ on the common apparent horizon thus provide us with a more concrete manifestation of the abstract remnant spins discussed earlier.

\begin{figure}[t,b]
\begin{overpic}[width=0.49\columnwidth]{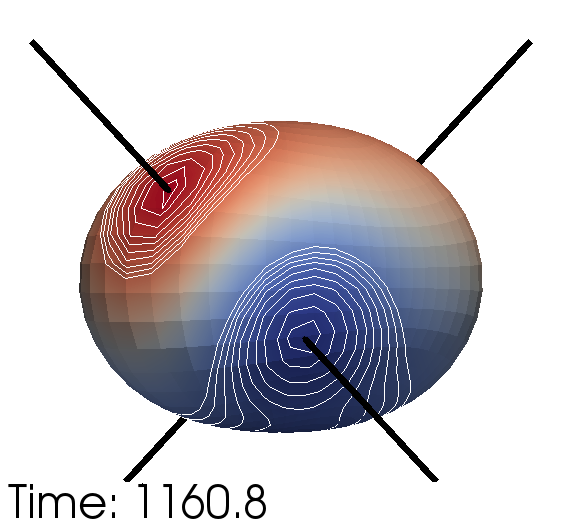}
\put(85,5){(a)}
\end{overpic}
\begin{overpic}[width=0.49\columnwidth]{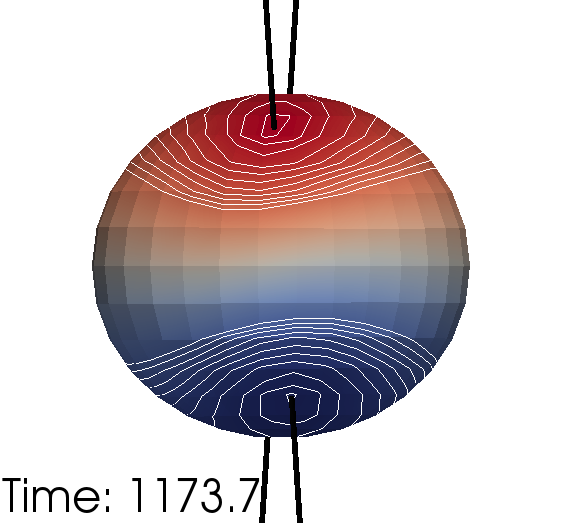}
\put(85,5){(b)}
\end{overpic}
\caption{The post-merger (during the merger phase) time evolution of the horizon vorticity $\mathcal{B}_{NN}$ for the SK simulation. The white curves are the contours of $\mathcal{B}_{NN}$. We have also drawn black lines connecting the centers of the spin patches for both remnant spins. Over time, the spin patches become more aligned with the $\bm{J}^{\rm tot}$ direction.
}
\label{fig:EvolutionOfContourPostMerger2}
\end{figure}

We can now use the spin patches to track the pre- and post-merger (remnant) spin dynamics. First of all, we note that the SK- simulation is the same as the SK simulation aside from a reversal of the individual spins. Since Eq.~\ref{eq:SpinSpin1PN} is invariant under $\bm{S_1} \rightarrow -\bm{S}_1$ and $\bm{S}_2 \rightarrow -\bm{S}_2$, we should observe the spins lifting up into the same side of the orbital plane for these two simulations \footnote{Provided that the black holes merge at similar orbital configurations for the two simulations, which appears to be the case. For example, the merger time is the same as shown in the time labels of Fig.~\ref{fig:EvolutionOfContourPostMerger2} (c)/(d) and (e)/(f). }. A comparison between Fig.~\ref{fig:EvolutionOfContourPostMerger2} (c)/(d) and (e)/(f) confirms this expectation, and provides us with some confidence that the spin-spin coupling is indeed responsible for generating the $S^z_1$ and $S^z_2$ components at merger. 

Using the numerical values for the spins (measured essentially with Eq.~\ref{eq:HorSpin1} on each individual apparent horizon) and the numerical trajectories of the black holes in the simulation coordinates, we can further make a prediction for the 
$z$ components of the spins by integrating Eqs.~\eqref{eq:SpinSpin1PN} and \eqref{eq:SpinOrbit}. 
Namely, we calculate the $\dot{S}^z_1$ values at a dense collection of times for one of the spins according to Eqs.~\eqref{eq:SpinSpin1PN} and \eqref{eq:SpinOrbit}, using the numerically measured $\bm{S}_1$, $\bm{S}_2$ and $\bm{n}$, before adding these $\dot{S}^z_1$ increments up into a predicted history for $S^z_1$.  
The results are shown in Fig.~\ref{fig:AnalyticalSpin}. We see a steep rise of $S_1^z$ for the SK, SK- and SK$\bot$ simulations towards merger, matching our $\mathcal{B}_{NN}$ spin patch observations. In particular, the prediction is for a dimensionless spin with $S^z_1 \approx 0.2$ just before the merger for all three simulations, which translates into an angle of $\sin^{-1}(0.2/0.5)\approx 0.4$ that $\bm{S}_1$ spans with the orbital plane. 
Taking the SK$\bot$ simulation for example, the line connecting the centers of the $+$ve $\mathcal{B}_{NN}$ and $-$ve $\mathcal{B}_{NN}$ spin patches for this simulation (see Fig.~\ref{fig:HorizonVorticity} (g) and (h)) spans an angle of $0.368$ with the orbital plane, which is fairly close to the prediction. The closeness between these two numbers is somewhat surprising, in that the simulation gauge is not the same as the harmonic gauge used for the PN calculations, and that we are operating in a regime close to merger. To further test the quality of the PN prediction, we produce the predicted histories for $S^x_1$ and $S^y_1$ using the same prescription, and compare them to their numerically computed counterparts in Fig.~\ref{fig:AnalyticalSpin}, which also show general agreement. 

The situation is different when we compare the predicted and the numerically computed $S^z_1$, as the latter lacks the steep rise just before merger that is also seen in spin patches. To understand this, recall that the numerical spin values are calculated as integrals that are essentially Eq.~\eqref{eq:HorSpin1}. 
Such an integration over the entire horizon does not distinguish the spin patches like $C$ and $G$ in Fig.~\ref{fig:HorizonVorticity} (c) and (e) from the mass quadrupole induced patches $A$ and $B$ as in Fig.~\ref{fig:HorizonVorticity} (c). As the mass quadrupole patches on each apparent horizon resembles a spin pointing in the $-\bm{\hat{z}}$ direction at merger time (see Fig.~\ref{fig:HorizonVorticity} (c), (e), (g) and (h)), the spin measurement $S^z_1$ according to Eq.~\eqref{eq:HorSpin1} could be negative even when the actual $S^z_1$ is positive. On the other hand, the measurements on $S^x_1$ and $S^y_1$ are less affected by this contamination. 
Furthermore, it is required that the first derivative of the scalar $\zeta$ in Eq.~\eqref{eq:HorSpin1} should be a rotation generating approximate Killing vector \cite{Dreyer2003,Cook2007, OwenThesis, Lovelace2008}, which may not be applicable when we approach the highly non-stationary merger phase. 

One interesting feature we have observed with our simulations is that at the merger, the spins seem to have been preferentially lifted out of the orbital plane towards the spin-$\bm{J}^{\rm tot}$ alignment side (see Fig.~\ref{fig:AnalyticalSpin} (d)). This trend also continues post-merger, as shown in Fig.~\ref{fig:EvolutionOfContourPostMerger2} for the SK simulation (similar behavior is observed for SK- and SK$\bot$), until a spin-$\bm{J}^{\rm tot}$ near-alignment is achieved, in agreement with our proposal in Sec.~\ref{sec:ExitStrategy}. The SK$\bot$ simulation is particularly interesting in that its initial spin orientations are chosen such that if the black holes merge at exactly the same orbital configuration as the SK and SK- cases, we would have $\dot{S}^z_1 < 0$ at merger according to Eq.~\eqref{eq:SpinSpin1PN}. Instead, the black holes merge with $\dot{S}^z_1 > 0$ (see Fig.~\ref{fig:HorizonVorticity} (g), (h) and Fig.~\ref{fig:AnalyticalSpin} (c), (d)) at an earlier time.
Therefore, based on our small sample of simulations, there appears to be a preference towards spin-$\bm{J}^{\rm tot}$ alignment at merger, which is not reversed post-merger. The exceptional case is the AA simulation, where the spins remain anti-aligned with $\bm{J}^{\rm tot}$ throughout the inspiral and merger phases, perhaps because the spins are stuck in a (possibly unstable) equilibrium configuration.

One possible explanation for the preference at merger time is provided by the spins' influence on the orbital motion, through the extra acceleration \cite{Barker:1975ae,1979GReGr..11..149B,1993PhRvD..47.4183K,Kidder:1995zr,Thorne1985,Thorne:1987af,BrumbergBook} 
\bea \label{eq:acc}
\bm{a} &=&  -\frac{3}{\mu R^4}\left[ \bm{n} (\bm{S}_1\cdot\bm{S}_2 ) + \bm{S}_1 (\bm{n}\cdot\bm{S}_2) + \bm{S}_2 (\bm{n}\cdot\bm{S}_1) \right. \notag \\
&& \left. - 5 \bm{n}(\bm{n}\cdot\bm{S}_1)(\bm{n}\cdot\bm{S}_2) \right]
\eea  
they impose on the relative one body equivalence to the two body motion. In Eq.~\eqref{eq:acc}, we have only shown the spin-spin contribution, as we are interested in the beginning of the final ascend of $S_1^z$ and $S_2^z$, and the spin-orbit contribution is small for our $\pi$-symmetric simulations when $S^z_1$ and $S^z_2$ are still small (i.e. $\bm{S}_1+\bm{S}_2 \approx 0$). It is plausible that the BBH coalescence progresses quickly towards merger when ${a}_n \equiv \bm{a} \cdot \bm{n} < 0$ (radially pulling the black holes closer) and ${a}_t \equiv \bm{a} \cdot (\bm{\hat{z}} \times \bm{n}) < 0$ (slowing down the transverse orbital motion of the black holes), such that the instantaneous impact parameter is altered in the direction conducive to merger. 
We note that only the second and third terms in the square bracket of Eq.~\eqref{eq:acc} contribute to $a_t$, and these two terms have the same value in our $\bm{S}_1 \approx -\bm{S}_2$ context. Therefore we have 
\bea 
a_t &=&  -\frac{6}{\mu R^4} \left[(\bm{\hat{z}} \times \bm{n}) \cdot \bm{S}_1\right] (\bm{n}\cdot\bm{S}_2) \notag \\
&=& -\frac{6}{\mu R^4}\bm{\hat{z}} \cdot \left[(\bm{n}\cdot\bm{S}_2) \bm{n}\times \bm{S}_1\right].\label{eq:AtExp}
\eea
Comparing Eq.~\eqref{eq:AtExp} with Eq.~\eqref{eq:SpinSpin1PN}, we arrive at 
\bea
a_t = -\frac{2}{\mu R} \dot{S}^z_1,
\eea
so that regions of $a_t < 0$ always correspond to $\dot{S}^z_1 > 0$, and there would subsequently be a statistical preference for spin-$\bm{J}^{\rm tot}$ alignment at merger.  

\section{Discussion \label{sec:Discussion}}
We note that an observation in this paper may help achieve a consistency between the QNM frequencies. 
First recall that an electric/magnetic parity quasinormal mode is defined to be one whose corresponding metric perturbation has the parity matching the sign of $(-1)^l$/$(-1)^{l+1}$. Because the $\bm{\mathcal{B}}$/$\bm{\mathcal{E}}$ field has the opposite/same parity to the underlying metric perturbations \cite{Nichols:2012jn}, the mass/current quadrupole would then excite the electric/magnetic parity $l=2$ QNMs (see also Ref.~\cite{Nichols:2011ih}, as well as Ref.~\cite{Nichols:2012jn} for discussions on the similarities between the mass/current quadrupole generated $\bm{\mathcal{E}}$ and $\bm{\mathcal{B}}$ fields and those associated with the electric/magnetic parity QNMs). The parity properties of the various quantities are summarized in Table ~\ref{tb:Parities}.

The electric and magnetic parity QNMs are degenerate in that they share the same complex frequency (see e.g. Sec.~IC3 of Ref.~\cite{Nichols:2012jn} and Sec.~VB1 of Ref.~\cite{Nichols:2011ih}). This has an interesting consequence, as was noted by Ref.~\cite{Nichols:2011ih} when motivating spin-locking in the superkick configurations. Namely, when QNMs of both parities are present, the current quadrupole should evolve at the same frequency as the mass quadrupole at the end of merger (just before the onset of the QNM ringdown phase). For the superkick configurations, because the mass quadrupole evolves at twice the orbital frequency, while the current quadrupole's frequency is essentially a sum of the orbital and the spin precession frequencies, this further implies that the spin precession frequency must lock onto the orbital frequency \cite{Nichols:2011ih}.

\begin{table}[!b]
\begin{tabular}{c|c|c|c|c}
  \hline
Quad. Moment
& \hspace{1mm} $\bm{\mathcal{E}}$ \hspace{1mm}
& \hspace{1mm} $\bm{\mathcal{B}}$ \hspace{1mm}
& Metric Pert.
& Excited $l=2$ QNMs
\\ \hline
\hline
Mass & + & - & + & Electric
\\ \hline
Current & - & + & - & Magnetic
\\ \hline
\end{tabular}
\caption{The parity of various quantities generated by the mass and current quadrupole moments. }
\label{tb:Parities}
\end{table}

A robust mechanism must be present for this locking to occur. A calculation using the leading order PN spin-orbit coupling expression \eqref{eq:SpinOrbit} for a $\pi-$symmetric superkick configuration yields \cite{Nichols:2011ih,Bruegmann-Gonzalez-Hannam-etal:2007}
\bea \label{eq:SpinPrecession1}
\dot{\beta}(t) = \frac{7M}{8R(t)}\Omega(t),
\eea
where $\dot{\beta}$ is the spin precession frequency. Therefore, as the black holes move closer to each other, $\dot{\beta}$ can approach $\Omega$, and with a mixed use of gauge \cite{Nichols:2011ih}, equalize with it \cite{Bruegmann-Gonzalez-Hannam-etal:2007}. However, this equality is broken again when $R(t)$ reduces further. So instead of locking, we have only a momentary coincidence.
Another mechanism for locking is proposed by Ref.~\cite{Nichols:2011ih}, which considers geodetic precession in black hole perturbation theory.  This alternative provides a stronger precession
\bea \label{eq:SpinPrecession2}
\dot{\beta}(t) = \frac{3M}{R(t)}\Omega(t),
\eea
but is otherwise similar to the leading order PN spin-orbit coupling result. Without invoking further dynamics, one would then be forced to make the inference that the BBH QNM ringdown begins, and that the spacetime dynamics that can be construed as two individual black holes approaching each other, ceases, precisely at the $R(t)$ that gives $\dot{\beta}=\Omega$. Note that Eqs.~\eqref{eq:SpinPrecession1} and \eqref{eq:SpinPrecession2} do not depend on the magnitude of the spins, so infinitesimal spins would appear to still play a vital role in the transition into the QNM ringdown phase. So additional dynamics are likely involved. For example, if the magnetic parity QNMs are completely absent, so that their frequencies become irrelevant, then the consistency would be achieved by default.  
Perhaps more interestingly, as discussed in relation to the next-to-leading-order PN corrections to $U_{\text{SO}}$ in Sec.~\ref{sec:ExitStrategy}, the energetically favorable orientation of the spins have a frame-dependent offset from the directions of $\pm \bm{J}^{\rm tot}$. As this offset can evolve at the frequency of $\Omega$ when the precession of $\bm{S}_1$ is locked onto the orbital motion, we could in principle have a self-consistent sustained locking (as opposed to a momentary coincidence) scenario if the spins are kept in these orientations by, for example, the same dynamics that drove them there in the first place. In other words, the mechanism responsible for the decline of the current quadrupole moment may also be responsible for locking its frequency to the desired value. In this case, the magnetic parity QNMs will not need to vanish in the ringdown waveform. However, we are operating in a regime where the PN expressions are not expected to remain fully valid. To see whether similar effects are actually present in a strong field and fast motion setting, we plan to carry out further studies at a later date. 

\acknowledgments
The author is grateful to Robert Owen for carrying out a previous incarnation of the SK simulation. 
We thank David Nichols, Sean McWilliams, Aaron Zimmerman and Robert Owen for carefully reading a draft of the paper, and providing valuable suggestions. 
We would also like to thank Yanbei Chen, Eliu Huerta, Lawrence Kidder, Geoffrey Lovelace, Harald Pfeiffer, Mark Scheel and Kip Thorne 
for useful discussions and help on references. The simulations and visualizations in this work are performed 
on the WVU computer clusters \textsc{Spruce Knob} and \textsc{Mountaineer}, and the Caltech cluster \textsc{Zwicky}
funded by the Sherman Fairchild Foundation and the NSF MRI-R2 grant PHY-0960291.
\appendix

\bibliography{paper.bbl}

\end{document}